\documentstyle[preprint,aps]{revtex}
\begin{document} % 14/12/93
\preprint{DFPD 93/TH/78}
\title{Complex Conjugate Pairs in Stationary
Sturmians}
\author{P. J. Dortmans, L. Canton and G. Pisent}
\address{Istituto Nazionale di Fisica Nucleare e Dipartimento
di Fisica dell'Universit\`{a} di Padova,\\
 via Marzolo 8, Padova I-35131,
Italia}
\author{K. Amos}
\address{School of Physics, University of Melbourne, Parkville, 3052
Victoria Australia}

\maketitle

\begin{abstract}
Sturmian eigenstates specified by  stationary scattering
boundary conditions are particularly useful in  contexts
such as forming simple separable two nucleon t matrices,
and are determined via solution of
generalised eigenvalue equation using real
and symmetric matrices. In general, the spectrum of such an equation
may contain complex eigenvalues. But to each
complex eigenvalue there is a corresponding conjugate partner.
In studies using realistic nucleon--nucleon potentials, and in
certain positive energy intervals, these complex conjugated pairs
indeed appear in the Sturmian spectrum. However,
as we demonstrate herein, it is  possible to recombine the complex
conjugate pairs and corresponding states into a new, sign--definite
pair of real quantities with
which to effect separable expansions of the (real) nucleon--nucleon
reactance matrices.
\end{abstract}


Sturmian states are solutions of Schr\"{o}dinger--like equations in which
the energy is treated as a continuous parameter
and the strength of the potential plays
the role of the spectral variable \cite{Wein63}.
Although at positive energies often one solves \cite{Rawe82} the
Sturmian equations with outgoing
boundary conditions (thereby obtaining eigenvalues and states
which are complex), it is convenient sometimes to evaluate the  Sturmians
under stationary boundary conditions \cite{Tani65}.
Those states are particularly useful in forming separable expansions
of the nucleon--nucleon (NN) reactance ($K$) matrices, and the process
of so doing defines the Sturmian splitting method (SSM) \cite{Pise93}.
Therein, the stationary Sturmian expansion was used to separate those
$K$ matrices into two groups. Each group is a sign definite
separable representation in itself, one set having attractive and the
other having repulsive character. In fact, a set of separable
$K$ matrices was generated and  the Heitler equation
used to specify the $t$ matrices thereafter.  The process however allows
us to avoid any pathology in the related $t$ matrices due to vanishing
on--shell phase shifts (or the equivalent coupled channel relation).
Such pathologies do
occur \cite{Cant90,Dort91} when  other similar separable expansion
methods \cite{Bart86} are used with realistic (NN) interactions.

To evaluate stationary Sturmians, we solve a
generalised eigenvalue equation with purely real, symmetric matrices.
Naively one might believe that the corresponding (stationary)
eigenvalues and states would also be real. But that is not necessarily
the case. If the potential term in the eigenvalue equation
has both an attractive and repulsive character,
as with realistic NN interactions,
degeneracies in the eigenvalues lead to some of them becoming complex.
However, those eigenvalues always occur in
complex conjugate pairs (CCPs) and, with a simple
recombination of such eigenvalues and states, equivalent pairs
can be defined that are purely real and have opposite signs.
With them, all contributions to the resultant sum
in the $K$--matrix expansion are real and sign definite.

We give now a basic outline of how the Sturmian
eigenvalues are determined, and describe why and where these CCPs occur.

To evaluate  Sturmians, ($\phi^{(P)}_{l,s}(q;E)$),
under stationary boundary conditions,
 in momentum  space
one must solve the generalised eigenvalue problem \cite{Cata91},
\begin{equation}
\sum_l\int^{\infty}_{0}
U^{(P)}_{Ll}(p,q;E)\phi^{(P)}_{l,s}(q;E) q^2dq     =
\eta^{(P)}_{s}(E)\sum_l \int^{\infty}_{0}
V_{Ll}(p,q) \phi^{(P)}_{l,s}(q;E) q^2dq     \ ,\label{a1}
\end{equation}
where $V_{ll'}(p,q)$ is the momentum space
interaction, and $U_{LL'}^{(P)}(p,q;E)$ is given by the principal value
of the second order Born term, i.e.
\begin{equation}
U^{(P)}_{LL'}(p,p';E) = \sum_{l}
{\bf P}\int^{\infty}_{0}{1\over E-q^2}
V_{Ll}(p,q)V_{lL'}(q,p')q^2dq     \ .
\label{Udef}
\end{equation}
The superscript $(P)$ denotes use of the stationary
boundary conditions.
It is convenient to define the Sturmian expansion via
\begin{equation}
\chi_{L,s}(p';E) = \sum_l \int^{\infty}_{0}
V_{Ll}(p',q) \phi_{l,s}(q;E) q^2dq\ , \label{KKstur}
\end{equation}
in terms of which  the (real) $K$ matrices are specified by \cite{Pise93}
\begin{equation}
K_{LL'}(p',p;E) = -\sum_{s=1}^\infty \chi_{L,s}^{(P)}(p';E)
\left\{{1\over \eta^{(P)}_s(E)(1-\eta^{(P)}_s(E))}\right\}
\chi_{L',s}^{(P)}(p;E)\ .\label{Kstur}
\end{equation}
As the $K$ matrices are real,
one might expect the eigenvalues ($\eta$) and states ($\chi$)
to be real also. But that is not always the case.
Indeed for positive energies, eigenvalues that are
CCPs occur whenever the interaction
produces degenerate eigenvalues.
Further development is needed to ensure that when such occur,
the separable expansions of the $K$ matrices remain real.

We consider first a special interaction
studied previously \cite{Cant90}, and with which
the stationary Sturmians are analytic.
In this case the states $\chi_{s}(p;E)$ may be written as
\begin{equation}
\chi_{s}(p;E) = t^{-1}_{11} a_{1,s}(E) u_{1}(p) +
t^{-1}_{22} a_{2,s}(E) u_{2}(p)\ ,
\end{equation}
where the two components, $t_{ii}$,
\cite{Cant90} have differing signs,
and the two energy dependent coefficients
$a_{i,s}$ are solutions of the real non symmetric standard 2 $\times$ 2
eigenvalue problem
\begin{equation}
\sum_{j=1,2} {\bf G}_{i,j}^{(P)} t^{-1}_{jj} a_{j,s} = \eta^{(P)}_{s}
a_{i,s}\ .
\end{equation}
Here ${\bf G}_{i,j}^{(P)} = t_{ii}\delta_{ij} - P_{ij}$, where
$P_{ij}$ is given by equation (3.2) of ref.\cite{Cant90}.

The result is that a complex conjugated pair appears in the spectrum
when the on--shell momentum ($k$) is in the range (0.26--0.29) fm$^{-1}$.
Outside of that range, and for E $<$ 0, the two eigenvalues are real.
In this problem, the two eigenvalues are the roots of a binomial equation,
and one finds that at the extremities of this interval,
the two real eigenvalues are degenerate. From this,
one may observe that CCPs
result only if the potential has both an attractive and a repulsive
character.

We now consider just how the generalised
eigenvalue equations can be solved numerically.
The integrals in the semi--infinite interval $[0,\infty]$ of
Eq.(\ref{a1})
are found \cite{Cata91} using a standard Gauss--Laguerre
N--point quadrature formula with which this equation is transformed into a
generalised $2N\times 2N$ matrix eigenvalue problem
\begin{equation}
\sum_{j=1}^{2N}{\bf U}_{i,j}^{(P)} a_{j,s}
=\eta^{(P)}_s
\sum_{j=1}^{2N}{\bf V}_{i,j} a_{j,s}\ ,
\label{a2}
\end{equation}
where
\begin{eqnarray}
a_{j,s} &=& k_j^2 w_j\phi^{(P)}_{0,s}
(k_j,E) \hspace*{4.2cm} {\rm if}\  j\leq N\nonumber\\
a_{j,s} &=& k_{(j-N)}^2 w_{(j-N)}
\phi^{(P)}_{2,s}(k_{(j-N)},E) \hspace*{2.0cm}{\rm if}\  j > N\ ,
\end{eqnarray}
and $k_j$ and $w_j$ are the points and weights of the quadrature formula.

The matrices ${\bf U}^{(P)}$ and ${\bf V}$ are real and symmetric and so
if ${\bf V}$ is non singular and positive (or negative) definite, the
generalised eigenvalues are real and finite. But for realistic
NN interactions, the matrix  ${\bf V}$ is not sign definite
and therefore the generalised spectrum of Eq.(\ref{a2}) may then
contain CCPs of eigenvalues (and associated eigenvectors),
in agreement with the general theorem for generalised eigenvalue
problems \cite{Golu83}.

The same equation holds for negative energies
where stationary Sturmians now coincide with Weinberg's states.
These eigenvalues are known to be real in spite of the fact
that ${\bf V}$ is not sign definite since it can be shown that the states
$|\chi>$  satisfy an equivalent generalised equation of the type,
\begin{equation}
{\bf G}(E){\bf VG}(E)\ \vert\chi_s>=\eta_s(E){\bf G}(E)\ \vert\chi_s>\ ,
\end{equation}
and in which, for $E<0$, no singularity occurs. Both matrices are
real and symmetric and for $E<0$, ${\bf G}(E)$ is clearly
negative definite. Therefore the eigenvalues are real.

As a second example we considered the positive energy,
generalised stationary spectrum for the $^1S_0$ Reid soft
core potential \cite{Reid68}.
At 100 and 200 MeV all the eigenvalues are real.
But at 140 MeV, a CCP appears having the value $(-0.33\pm i 0.02)$ fm$^{-1}$.
It is a quite stable result,
changing little in value with variation either of the set of grid points or
of the energy (around 140 MeV). That CCP disappears between 189
and 190 MeV being replaced (at 190 MeV) by two real eigenvalues.
Those two real eigenvalues are almost degenerate
and are comparable
with the real part of the 189 MeV CCP. Thus
the CCP has not originated from round off errors
due to numerical approximations.
Rather it is an actual characteristic of the spectrum.

As indicated previously, we seek a technique to eliminate use
of these CCPs in, for example, the specification of separable representations
of NN operators using the SSM scheme. This is achieved by first grouping
those CCPs into
an attractive ($+$) and a repulsive ($-$) subspace according
to the following criteria. Given that we can
define the total contribution of a CCP in the $K$-matrix expansion as
\begin{equation}
K_{LL'}^{ccp}(p,p';i)=-[\chi^{(P)}_{L,i}(p) \mu_i
\chi^{(P)}_{L',i}(p') +
\{\chi^{(P)}_{L,i}(p) \mu_i \chi^{(P)}_{L',i}(p')\}^*]\ ,\label{KCCP}
\end{equation}
where
\begin{equation}
\mu_i\equiv {1\over [\eta^{(P)}_i(1-\eta^{(P)}_i)]}\ .
\end{equation}
We can also define the eigenvalue as
\begin{equation}
\mu_i\equiv \sigma_i+i\tau_i\ ,
\end{equation}
and the eigenstate as
\begin{equation}
\chi^{(P)}_{L,i}(p) \equiv f_{L,i}(p) + i g_{L,i}(p)\ .
\end{equation}
Now defining a new pair of eigenvalues with opposite sign via
\begin{equation}
\mu_i^\pm\ \equiv
\pm {2\over \left\{ \vert\eta^{(P)}_i(1-\eta^{(P)}_i)\vert\right\}}\ ,
\end{equation}
we obtain two purely real quantities, namely
\begin{eqnarray}
\mu^+_i &=& 2 \sqrt{\sigma_i^2+\tau_i^2}\nonumber\\
\mu^-_i &=& -2 \sqrt{\sigma_i^2+\tau_i^2}\ .
\end{eqnarray}
Similarly, defining the attractive and repulsive eigenstates as
\begin{eqnarray}
\chi_{L,i}^+(p) &=& a^+_{1}f_{L,i}(p) + a^+_{2}g_{L,i}(p)\nonumber\\
\chi_{L,i}^-(p) &=& a^-_{1}f_{L,i}(p) + a^-_{2}g_{L,i}(p)\ ,
\end{eqnarray}
the `a' coefficients are also purely real and are specified by
\begin{eqnarray}
a^+_{1} &=& {1\over \sqrt{2}}
{\sqrt{\sqrt{ \sigma_i^2 +\tau_i^2}+\sigma_i}\over
\ ^4\sqrt{ \sigma_i^2 +\tau_i^2}}
\nonumber\\
a^+_{2} &=& {-\tau_i\over \sqrt{2}|\tau_i|}
{\sqrt{\sqrt{ \sigma_i^2 +\tau_i^2}-\sigma_i}\over
\ ^4\sqrt{ \sigma_i^2 +\tau_i^2}}
\nonumber\\
a^-_{1} &=& {1\over \sqrt{2}}
{\sqrt{\sqrt{ \sigma_i^2 +\tau_i^2}-\sigma_i}\over
\ ^4\sqrt{ \sigma_i^2 +\tau_i^2}}
\nonumber\\
a^-_{2} &=& {\tau_i\over \sqrt{2}|\tau_i|}
{\sqrt{\sqrt{ \sigma_i^2 +\tau_i^2}+\sigma_i}\over
\ ^4\sqrt{ \sigma_i^2 +\tau_i^2}}\ .
\end{eqnarray}
It is  possible then to redefine Eq.(\ref{KCCP}) as
\begin{equation}
K_{L,L'}^{ccp}(p,p';i)=-\{\chi^+_{L,i}(p) \mu^+_i
\chi^+_{L',i}(p')
+ \chi^-_{L,i}(p) \mu^-_i\chi^-_{L',i}(p') \}\ ,
\end{equation}
where the term $K_{L,L'}^{ccp}(p,p';i)$ splits into two contributions
in which each element of the CCP contributes an equal amount to the
attractive and the repulsive quantities.

In summary, we have shown that CCPs may appear in the stationary
Sturmian spectrum for certain energy range. When this is the case,
the Sturmian eigenstates become degenerate at the edge of these
intervals leading to the transition from a real to a complex
eigenspectrum. The occurrence of such CCPs and corresponding eigenstates
leads to terms in the Sturmian expansion which are neither real nor
sign--definite; a coincidence which does not allow straightforward
application of the SSM. However,  a proper
recombination of the two complex conjugate states leads to new ones
which are real and sign--definite and with which application
of the SSM in the presence of CCPs can be made.

\end{document}